\documentclass[aip,amsmath,amssymb,reprint]{revtex4-2} 
\usepackage{color}
\usepackage{hyperref}
\usepackage{enumitem}
\usepackage{graphicx,subfigure}
\usepackage[normalem]{ulem}
\usepackage{lineno}
\usepackage{booktabs}
   \usepackage{array}
    \usepackage{multirow}
    \usepackage{hhline}
    \usepackage{makecell}
     \usepackage[symbol]{footmisc}
     \usepackage{balance}
     \usepackage{ulem}
\usepackage{xcolor}

\newcommand\redout{\bgroup\markoverwith
{\textcolor{red}{\rule[0.5ex]{2pt}{2pt}}}\ULon}
    \usepackage{siunitx} 
\newif\ifHighlitedChanges
\def\ifHighlitedChanges{\iftrue}
\ifHighlitedChanges
   
  \def\STRIKE#1{{\color{blue}\sout{#1}}}
\else
  
  \def\STRIKE#1{\relax}
\fi

\begin{document}
\title{Manifestations of the possible thermodynamic origin of water's anomalies in non-classical vapor nucleation at negative pressures}
\author{Yuvraj Singh}
\affiliation{Department of Physics, Indian Institute of Science Education and Research (IISER) Tirupati, Tirupati, Andhra Pradesh, 517619, India}
\author{Mantu Santra}
\affiliation{School of Chemical and Materials Sciences, Indian Institute of Technology Goa, Goa, 403401, India}
\author{Rakesh S. Singh}
\email{rssingh@iisertirupati.ac.in}
\affiliation{Department of Chemistry, Indian Institute of Science Education and Research (IISER) Tirupati, Tirupati, Andhra Pradesh 517619, India}
\begin{abstract}
Over the years, various scenarios --- such as the stability-limit conjecture (SLC), two critical point (TCP), critical point-free (CPF), and singularity-free (SF) --- have been proposed to explain the thermodynamic origin of supercooled water’s anomalies. However, direct experimental validation is challenging due to the rapid phase transition from metastable water. In this study, we explored whether the phase transition pathways from metastable water provide insight into the thermodynamic origin of these anomalies. Using a classical density functional theory approach with realistic theoretical water models, we examined how different thermodynamic scenarios influence vapor nucleation kinetics at negative pressures. Our findings show significant variations in nucleation kinetics and mechanism during both isobaric and isochoric cooling. In the TCP scenario, the scaled nucleation barrier ($\beta \Delta\Omega_{\rm v}^*$) increases steadily during isobaric cooling, with a slight decrease near the Widom line at lower temperatures ($T$s). In contrast, the SF scenario shows a monotonic increase in the nucleation barrier. For the CPF scenario, we observed a non-classical mechanism, such as wetting-mediated nucleation (where the growing vapor nucleus is wetted by the intermediate low-density liquid phase) and the Ostwald's step rule at low temperatures. Isochoric cooling pathways also revealed notable differences in $T$-dependent $\beta \Delta\Omega_{\rm v}^*$ trends between the TCP and CPF scenarios. Overall, this study underscores the importance of analyzing phase transition kinetics and mechanism to understand the precise thermodynamic origin of supercooled water’s anomalies.
\end{abstract}
\maketitle

\section{\label{sec:level1} Introduction} 
Water exhibits remarkable deviations from the behavior of typical liquids~\cite{bagchi_book, pablo_nat_2006}. For example, the thermodynamic response functions, such as, isothermal compressibility ($\kappa_T$) and isobaric heat capacity ($C_P$), show a sharp (divergence-like) increase as the temperature decreases below the ice's melting point (\textit{i.e.}, in the supercooled state)~\cite{angell_1973, speedy_1976, kanno_angell_1979, angell_1982, Tombari_anomaly_1999}. These deviations, often referred as ``anomalies", are known to have important implications in many of the chemical and biological processes in nature~\cite{brini2017water, pohorille2012water, kontogeorgis2022water}. Over the years, many scenarios have been proposed to understand the thermodynamic origin of these anomalies~\cite{speedy1982stability,poole_1992, poole1994effect,sastry_sf_1996}. 
 
The divergence-like increase in $\kappa_T$ and $C_P$ observed in experiments on real water closely resembles the behavior of these response functions near a critical point or phase stability boundary, such as a spinodal. However, the nature of the boundary or singularity that water may be approaching remains unclear and subject to debate. An initial attempt to shed light on the anomalous behavior of these response functions was made by Speedy in $1982$~\cite{speedy1982stability}. His hypothesis, known as the stability-limit conjecture (SLC), posited that if the line corresponding to the temperature of maximum density (TMD) --- which has a negative slope in the positive pressure region of the $P-T$ phase diagram --- were to retain this negative slope under negative pressure conditions, it would eventually intersect the liquid spinodal (LS). At the point of intersection, it becomes thermodynamically necessary for the slope of the LS to flatten to zero. As the temperature decreases further, the LS may re-enter the positive pressure region of the phase diagram. Hence, the magnitudes of the thermodynamic response functions increase as isobaric cooling paths approach the LS at low temperatures. 

A decade after the proposal of the SLC, Poole et al.~\cite{poole_1992} introduced an alternative scenario to explain these anomalies. Through molecular dynamics simulations of the ST2 water~\cite{stillinger1974improved}, Poole et al. attributed the anomalous behavior of $\kappa_T$ and $C_P$ in real water to the presence of a liquid-liquid critical point (LLCP) in the deeply supercooled region. They hypothesized the existence of two distinct polymorphic forms of liquid water: high-density liquid (HDL) and low-density liquid (LDL). The liquid-liquid coexistence (LLC) line terminates at the LLCP~\cite{poole_1992}. The sharp increase in $\kappa_T$ and $C_P$ as temperature decreases is attributed to the influence of the LLCP and the associated Widom line (or, the locus of maxima in thermodynamic response functions). This interpretation is widely referred to as the two-critical-point (TCP) scenario. Recent computer simulation studies employing more realistic water models (including neural network-based models~\cite{pablo_prl_2022, molinero_pnas}) provide direct evidence for the existence of the LLCP~\cite{palmer_chem_rev_2018, pablo_nature_2014, pablo_science,singh_tip4p_2017, singh_tip4p_2016, zaragoza_JCP_2020}. 

Both of the aforementioned scenarios (SLC and TCP) implicitly assume that a singularity --- whether an LLCP or a re-entrant LS at positive pressure --- is necessary to explain the anomalous behavior of thermodynamic response functions in supercooled water. However, alternative explanations have also been proposed to understand the behavior of supercooled water and other tetrahedral liquids, such as SiO$_2$. These include the singularity-free (SF) scenario~\cite{sastry_sf_1996} and the critical-point-free (CPF) scenario~\cite{poole1994effect}. The CPF scenario can be viewed as a variant of the TCP scenario, where the LLC line extends into the highly negative pressure region and terminates at the LS. In the SF scenario, Sastry et al.~\cite{sastry_sf_1996} proposed that the negative slope of the TMD is sufficient to cause an increase in $\kappa_T$ upon cooling, without the need for any singularity. Over the years, however, despite numerous efforts dedicated to unraveling the origin of water's anomalies through various experiments, conclusive evidence supporting these proposed theoretical explanations remains elusive due to the spontaneous ice crystallization at deeply supercooled conditions.

 The seminal works by Poole et al.~\cite{poole1994effect} and Chitnelawong et al.~\cite{chitnelawong2019stability} examined the phase behavior of the ``extended van der Waals (EVDW)" model by systematically varying the strength of hydrogen bonds (H-bonds). They observed that, the LLCP transitions from stable to the metastable liquid (with respect to vapor) region of the phase plane as the H-bond strength decreases. With a further decrease in the strength of the H-bond, the LLC line extends deeper into the negative pressure region and meets the LS. In this context, the CPF scenario specifically refers to the disappearance of the LLCP. At lower H-bond strengths, the coexistence line between the LDL and HDL persists, but the LDL phase terminates at the LDL spinodal line rather than the LLCP, indicating the absence of the coexisting LDL and HDL phases beyond this point (see Fig.~\ref{fig:s1} in the Supplementary Material). Thus, the authors demonstrated that it is possible to reproduce TCP, and CPF (including SLC) scenarios within the the framework of the EVDW model.  

The approach adopted in Refs.~\emph{\citenum{poole1994effect}} and~\emph{\citenum{chitnelawong2019stability}} provides an opportunity to turn around the problem of the origin of water's anomalies in the sense that --- rather than restricting the system to undergo phase transition under deeply metastable conditions, study the phase transition pathways and their connections to the proposed thermodynamic scenarios. Such studies have the potential to distinguish between various scenarios and provide deeper insights into the underlying mechanism associated with the thermodynamic condition-dependent changes in the free energy surface that govern water's behavior. In our previous study, we utilized the microscopic phenomenological water model by Truskett et al.~\cite{pablo_jcp_1999} to analyze the vapor nucleation barrier variations on isobaric and isochoric cooling at negative pressures within a specific H-bond strength context which gives rise to the TCP scenario. This study was inspired by studies that suggest that the shape of the TMD line at negative pressures has the potential to provide important insights into the origin of water’s thermodynamic anomalies in the supercooled state~\cite{angell_sci, pablo_nature_2013, caupin2005liquid, caupin2006cavitation, frederic_natphys_2013, frederic_pnas_2014, frederic_pnas_2016, singh_tip4p_2017_2, frederic_jpcl_2017}. We found a striking relationship between the thermodynamic anomalies in the supercooled state and the kinetics of vapor and ice nucleation. 

The aim of the current study is to explore the vapor nucleaton kinetics and mechanism from negative pressure water for various thermodynamic scenarios proposed to elucidate the origin of water's anomalies. Here, we have employed models developed by Poole et al.~\cite{poole_1992} (referred as Model $1$) and Truskett et al.~\cite{pablo_jcp_1999} (referred as Model 2) to examine the vapor nucleation kinetics and mechanism along different paths in the phase plane. Our findings suggest that the vapor nucleation mechanism could be a valuable tool to probe the possible origin of water's anomalies in the supercooled state. 
\section{Method Details}\label{model}
\subsection{Theoretical water models} \label{model_details}
We have used two theoretical water models --- named Model $1$ and Model $2$  (discussed in the previous section). Model $1$ is a phenomenological theoretical model where we do not have an explicit geometric criterion to account for the local packing effects on water molecules' H-bond strength. The molar Helmholtz free energy for this model is given as, 
\begin{equation}
    F_m = F_{\rm VDW} + 2F_{\rm HB}
\end{equation}
where, $F_{\rm VDW}$ is the molar Helmholtz free energy for the van der Waals fluid and $F_{\rm HB}$ is the molar Helmholtz free energy describing the behavior of the H-bonds which is given as~\cite{poole1994effect}, 
\begin{equation}
   F_{\rm HB} = -RT\ln[\Omega+ \exp (-\epsilon_{\rm HB}/RT)].
\end{equation}
Here, there are $\Omega$ possible non-H-bonded states, each possessing zero energy, along with one H-bonded state with an energy $\epsilon_{\rm HB}$. In this model, one can adjust $\epsilon_{\rm HB}$ to generate different scenarios (see Fig.~\ref{fig:s1} in the Supplementary Material)~\cite{poole1994effect, chitnelawong2019stability}. The values of $\epsilon_{\rm HB}$ used to generate these scenarios are $-22$ kJ/mol, and $-16$ kJ/mol, respectively. One remarkable observation from this model is that, in the CPF scenario, the liquid line of the liquid-vapor coexistence (LVC) meets the LS at two points --- the conventional liquid-vapour critical point (LVCP), and near the HDL branch, well below the LVCP (see Fig.~\ref{fig:s1}B in the Supplementary Material, and Figs.~$6$ and $7$ in Ref.~\emph{\citenum{chitnelawong2019stability}}). Following Poole and coworkers, we refer to this additional point as the Speedy point~\cite{poole1994effect}. Typically, the LVC line in simple liquids meets its spinodal line at the LVCP where the density of the vapor and liquid is the same, and hence, the phases become indistinguishable. In contrast, at the Speedy point, the liquid and vapor densities are significantly different. As already reported in Refs.~\emph{\citenum{poole1994effect}} and~\emph{\citenum{chitnelawong2019stability}}, in the CPF scenario, it is the manner in which the liquid coexistence line terminates that causes the liquid spinodal to reenter into the positive pressure region at low temperatures, resulting in what is known as the SLC scenario. 
\begin{figure}
    \centering
    \includegraphics[width = .48\textwidth]{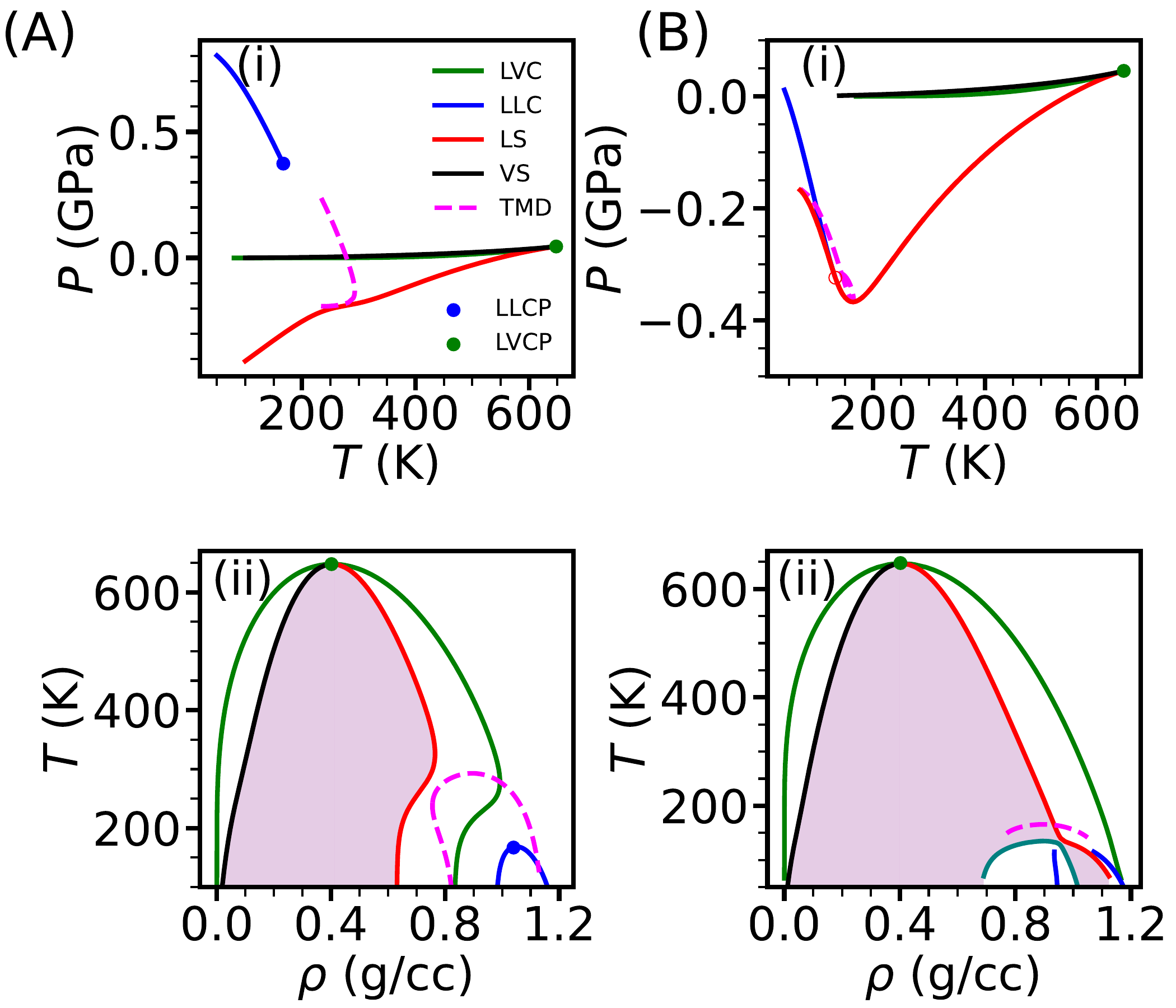}
    \caption{The $P-T$ (i) and $T-\rho$ (ii) phase diagram of Model $2$ representing the TCP (A), and CPF (B) scenarios. For the TCP scenario, the H-bond strength ($\epsilon_{\rm HB}$) is set to $-23$ kJ/mol, while for the CPF scenario, $\epsilon_{\rm HB} = -13$ kJ/mol. All other model parameters are same as in Ref.~\emph{\citenum{pablo_jcp_1999}}. Here, VS refers to the vapor spinodal. In the CPF scenario, the LLC line extends into the deeply negative pressure region and terminates at the LDL spinodal in the $P-T$ plane. Additionally, the LS line changes slope upon intersecting the TMD line and retraces towards the positive pressure region at lower temperatures. Therefore, similar to Model $1$, both TCP and CPF scenarios can be reproduced within this model by carefully tuning the H-bond strength.}
    \label{fig:1}
\end{figure}

The Model $2$, originally developed by Truskett et al.~\cite{pablo_jcp_1999}, is a microscopic phenomenological model which reproduces satisfactorily the phase behavior of liquid water. This model is geometrically tailored to emulate the behavior of water's H-bonds and incorporates various parameters regarding the range, strength, and orientation of these bonds. In this model, the interaction potential between the molecules ($\phi$) is decomposed into three different contributions: hard sphere ($\phi_{\rm HS}$), dispersion interaction ($\phi_{\rm DI}$), and H-bond ($\phi_{\rm HB}$), such that, $\phi = \phi_{\rm HS} + \phi_{\rm DI} + \phi_{\rm HB}$. The model includes a cavity with a radius of $r_i$ to mimic a low-density environment suitable for H-bonding.

Once the microscopic interaction potential is constructed, we can establish the relationship between thermodynamic quantities through the canonical partition function of the system~\cite{hill2013statistical}. 
 \begin{equation}
     Z(N,V,T) = \frac{1}{N! \Lambda^{3N}} \int \int d\textbf{r}^N d \Omega^N \exp(-\beta \phi),
 \end{equation}
 where $N$ is the number of particles, $V$ is the volume, $T$ is the temperature, $\Lambda$ is the thermal wavelength, $\beta = 1/k_{\rm B}T$, where $k_{\rm B}$ is the Boltzmann's constant. $\mathbf{r}$ and $\Omega$ represent the instantaneous molecular position and orientation, respectively. Considering these contributions and the geometric criteria for water's H-bonds, Truskett et al.~\cite{pablo_jcp_1999} derived the following expression for the canonical partition function:
 \begin{equation}
     Z(N,V,T) = \frac{1}{N! \Lambda^{3N}} (V-Nb)^N \exp(N\beta \rho a) (4\pi)^N \prod_{j=1}^8 f_j^{NP_j},
 \end{equation}
 where $b$ is the excluded volume per molecule, $\rho$ is the molecule number density ($N/V$), and $f_j$ is given by the following expression,
 \begin{equation}
     f_j = 1+ \frac{j}{4}(1-\cos(\phi^*))^2 (\exp\left[\beta \epsilon_j\right]-1).
 \end{equation}
In above expression, $P_j$ is related to the radial distribution function and the parameter $\phi^*$ regulates the proper orientation of molecules required for the H-bond formation. $\epsilon_j$ is associated with the crowding impact on water's potential energy, $\epsilon_j = \epsilon_{\rm HB} + (j - 1)\epsilon_{\rm p}$, where $j$ denotes the number of non-bonding molecules within the H-bond-forming shell,   $\epsilon_{\rm HB}$ and $\epsilon_{\rm p}$ are the H-bond strength and the penalty attributed to each non-bonding molecule within the shell, respectively. In the originally constructed model which exhibits the TCP scenario, $\epsilon_{\rm HB}$ is set to $-23$ kJ/mol with a penalty of $3$ kJ/mol for each non-bonded water molecule that crowds the central water molecule. Hence, the strength of the H-bond after addition of $j$ non-bonded molecules becomes $\epsilon_j = \epsilon_{\rm HB} + (j-1)\epsilon_{\rm p}$. This implies that there could be a maximum of $8$ non-bonding molecules of water within the H-bonding shell before the central molecule ceases to form any hydrogen bond (see Ref.~\emph{\citenum{pablo_jcp_1999}} for further details). Once the partition function is formulated, we can get the Helmholtz free energy as $F(N,V,T) = -k_{\rm B} T \ln(Z(N,V,T))$.   

\subsection{Phase behavior of Model $2$}\label{phase}
To generate the TCP and CPF scenarios for Model $2$, we systematically varied $\epsilon_{\rm HB}$ as well as $\epsilon_{\rm p}$, ensuring that a maximum of eight water molecules could form hydrogen bonds before the H-bond strength diminishes. This was based on the assumption that as $\epsilon_{\rm HB}$ decreases, the penalty for incorporating a single non-bonding water molecule within the hydration shell also decreases. In this study, we have only modified $\epsilon_{\rm HB}$ and adjusted $\epsilon_{\rm p}$ accordingly, while keeping all other model parameters unchanged from the original work that reproduces the TCP scenario. In Fig.~\ref{fig:1} we present two distinct scenarios: (a) the TCP scenario, where the LLCP resides within the stable liquid region with respect to the vapor (Figs.~\ref{fig:1}A(i) and ~\ref{fig:1}A(ii)), and (b) the CPF scenario. In the CPF scenario, the LLC line does not terminate at the LLCP; instead, it extends into the deeply negative pressure region and ends at the LDL spinodal (Figs.1B(i) and~1B(ii)). Furthermore, the LS line changes slope upon intersecting the TMD line and retraces towards the positive pressure region at lower temperatures. This behavior suggests that, similar to Model $2$, the CPF and SLC scenarios are interconnected in Model $1$ as well. The LS retraces towards the positive pressure region in the $P-T$ plane only for the CPF scenario. 

Although both models exhibit similar phase behavior, there are noteworthy differences between them. For instance, in the CPF scenario of Model $1$, the liquid branch of the LVC shows a slight hump and terminates at the LS at approximately $190$ K (see Fig.~S1 in the Supplementary Material). However, in Model $2$, it does not exhibit a hump and continues to extend downward at lower temperatures (see Fig.~\ref{fig:1}B(ii)). Furthermore, unlike Model $1$, the liquid branch of the LVC in Model $2$ encloses the LS in the $T-\rho$ plane (Fig.~\ref{fig:1}B(ii)). Therefore, even after the LS reverses its slope (positive to negative) in the $P-T$ plane, it does not re-enter to the positive pressure region of the phase diagram. However, we do not dismiss the possibility that, within the framework of Model $2$, the same phase behavior as that of Model $1$ could potentially be achieved by carefully choosing the model parameters. 
\section{Results and Discussion}
\subsection{Liquid-vapor interfacial surface energy}
The liquid-vapor interfacial free energy (or, surface tension, $\gamma_{\rm lv}$) is a crucial quantity that significantly influences nucleation kinetics. Therefore, it is important to first investigate the behavior of the surface tension between the coexisting liquid and vapor phases. Here, we estimated $\gamma_{\rm lv}$ along the LVC line for both water models across various $\epsilon_{\rm HB}$ values, each representing a different scenario. We used classical density functional theory (CDFT) to calculate the interfacial surface tension with a Cahn-Hilliard square-gradient free energy functional, that is given as~\cite{ch_1958, oxtoby_1998},
\begin{equation}
    \Omega[\rho(z)] = \int dz [f(\rho(z)) - \mu \rho(z)] + \frac{K_\rho}{2} \int dz  [\bigtriangledown \rho(z)^2],
    \label{E6}
\end{equation}
where $\Omega[\rho(z)]$ is the grand canonical free energy functional of the inhomogeneous system, $f$ is the Helmholtz free energy density ($F/V$), $\mu$ is the chemical potential, and $K_\rho$ is the density correlation length. 

We first obtained the equilibrium density profile $\rho(z)$ by minimizing the Euler-Lagrange equation, $\delta \Omega[\rho(z)] / \delta \rho(z) = 0$, and solved the resulting differential equation with appropriate boundary conditions ($\rho = \rho_{\rm v}$ at $z=0$, and $\rho = \rho_{\rm l}$ as $z \to \infty$, where $\rho_{\rm l}$ and $\rho_{\rm v}$ denote the coexistence densities of liquid and vapor phases, respectively, at a given temperature). After obtaining the density profile $\rho(z)$, we calculated $\gamma_{\rm lv}$ using $\gamma_{\rm lv} = (\Omega[\rho(z)] - \Omega[\rho_{\rm l}]) / A$, where $A$ is the area of the interface. For Model 2, we determined the correlation length $K_\rho$ microscopically from model parameters using $K_\rho = -\frac{\pi}{3} \int r^4 \phi(r) dr$, where $\phi(r)$ is the attractive part of the inter-particle interaction potential~\cite{Rowlinson1979TranslationOJ}. For Model $1$, however, since the interaction potential form is not explicitly modeled, we judiciously chose $K_\rho$ = $5.0$ (in reduced units) to ensure that $\gamma_{\rm lv}$ values obtained from CDFT are consistent in magnitude with experimental and simulation values for various water models~\cite{vega_2007} at same temperatures near ambient conditions.
\begin{figure}
    \centering
    \includegraphics[width = .48\textwidth]{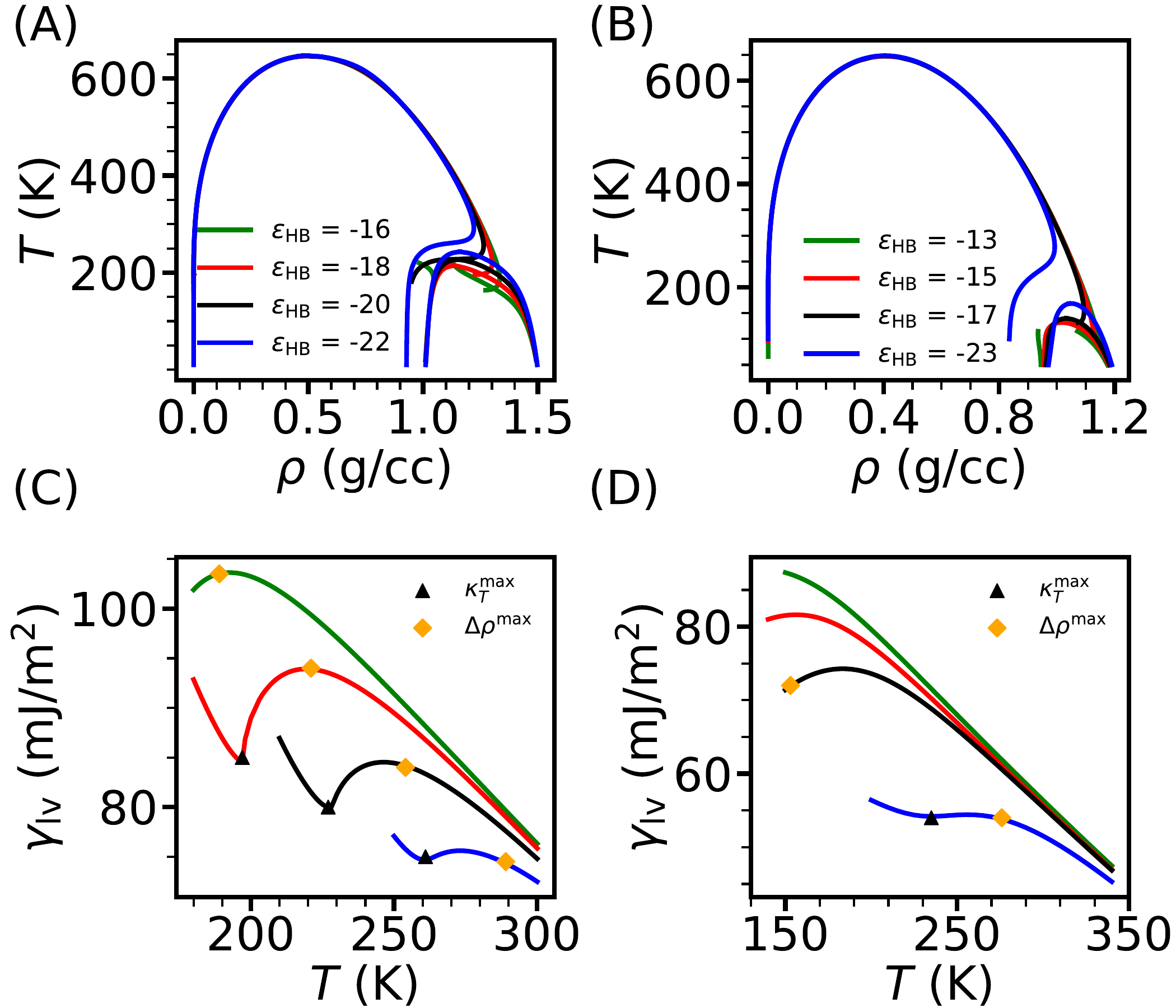}
    \caption{The $T-\rho$ phase diagram shows the LVC and LLC lines for Model $1$ (A) and Model $2$ (B) (see Fig.~\ref{fig:1} in Section~\ref{model} and Fig.~\ref{fig:s1} in the Supplementary Material). Here, we have varied $\epsilon_{\rm HB}$ (in units of kJ/mol) to realize different scenarios ranging from TCP to CPF. The vapor-liquid surface tension ($\gamma_{\rm lv}$) along the LVC is shown for Model $1$ (C) and Model $2$ (D). The temperature of maximum isothermal compressibility of liquid water ($\kappa_T^{\rm max}$) along the LVC line is marked with $\blacktriangle$, and $\blacklozenge$ denotes the temperature of maximum density difference between the coexisting phases ($\Delta \rho^{\rm max}$) along the LVC.}    
    \label{fig:2}
\end{figure}

Figures~\ref{fig:2}A and~\ref{fig:2}B depict the $T$-$\rho$ phase diagrams for Models 1 and 2, respectively, with different $\epsilon_{\rm HB}$ values corresponding to different scenarios ranging from TCP to CPF. For simplicity, we have shown only the LVC and liquid-liquid coexistence (LLC) curves. In Figs.~\ref{fig:2}C and\ref{fig:2}D, we present $\gamma_{\rm lv}$ along the LVC for both the models. For each $\epsilon_{\rm HB}$, $\gamma_{\rm lv}$ initially increases as temperature decreases, reaching a maximum, then a minimum, at lower temperatures. Notably, the minimum surface tension ($\gamma_{\rm lv}^{\rm min}$) occurs near the temperature of maximum $\kappa_T$ along the LVC. This non-monotonic $T$-dependent behavior of $\gamma_{\rm lv}$ has not been reported in previous computational studies~\cite{malek_2019_sip, wang_2019_sip, dellago_2022_sip, lu_2006_sip, rogers2016possible, vins_2020_sip}, although recent CDFT-based theoretical studies suggest similar non-monotonic trends in $\gamma_{\rm lv}$~\cite{mikhail_jpcb_2023, singh2023anomalous}. Below the temperature of maximum $\kappa_T$, $\gamma_{\rm lv}$ increases as $T$ decreases further. We also found that $\gamma_{\rm lv}$ increases as $\epsilon_{\rm HB}$ decreases, and this increase is more pronounced at lower temperatures. 

Interestingly, the temperature of maximum surface tension does not fully correlate with the temperature of maximum density difference between the coexisting phases along the LVC. This may result from factors, such as the shape and curvature of the free energy basins for the coexisting liquid and vapor phases, that influence $\gamma_{\rm lv}$. We have not shown here the surface tension data for the SF scenario, as this has been previously reported for Model 2 by Feeney and Debenedetti~\cite{debenedetti_sip}, who observed a second inflection point (SIP) in the $T$-dependent behavior of $\gamma_{\rm lv}$ at lower temperatures. Recent simulations~\cite{vins_2020_sip, wang_2019_sip, dellago_2022_sip, lu_2006_sip, mikhail_jpcb_2023} and experimental investigations~\cite{vins_2020_sip} of water's interfacial surface tension also suggest a SIP in the $T$-dependence of $\gamma_{\rm lv}$. 
\begin{figure*}
    \centering
    \includegraphics[width=0.96\textwidth]{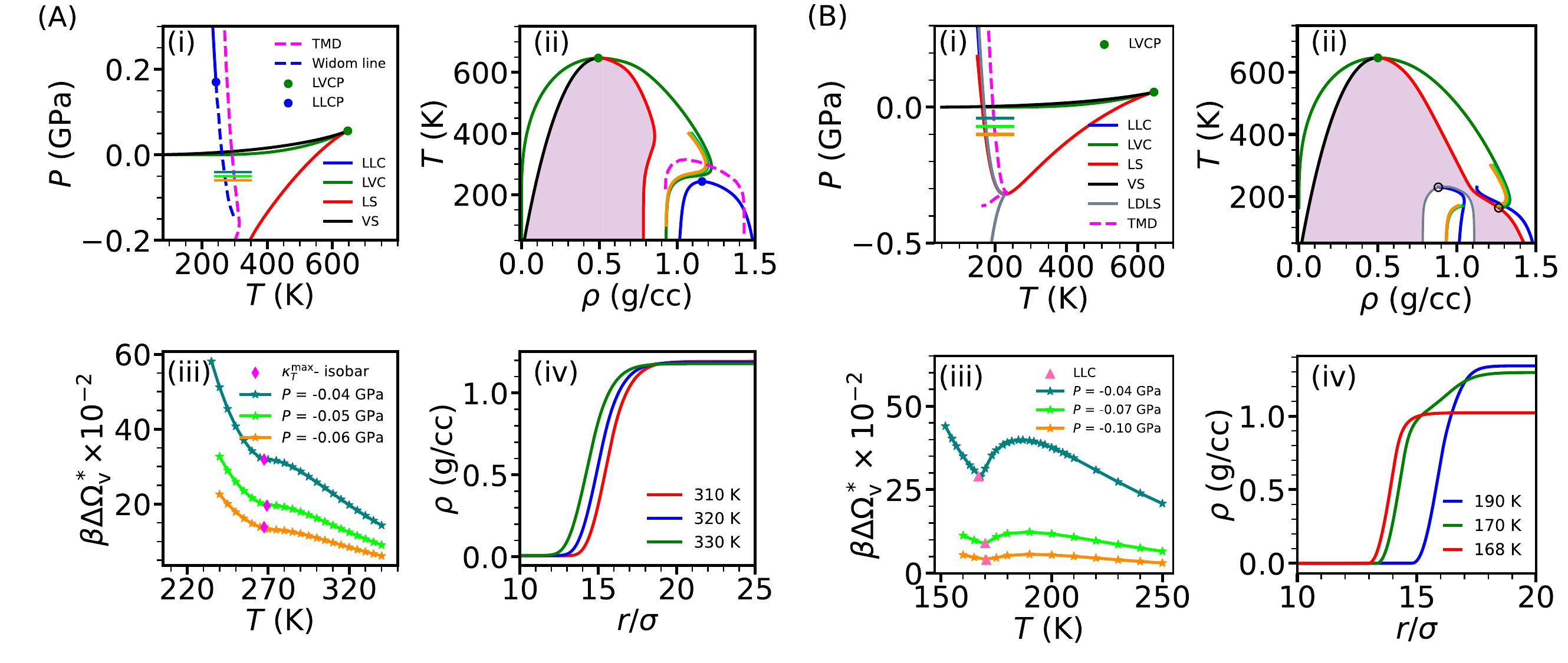}
    \caption{(A) The phase diagram for the TCP scenario of Model $1$ ($\epsilon_{\rm HB} = -22$ kJ/mol) is shown in the (i) $P-T$, and (ii) $T-\rho$ planes, along with the (iii) scaled vapor nucleation barrier ($\beta \Delta \Omega_{\rm v}^*$) for isobaric cooling paths highlighted in (i) and (ii). The $\blacklozenge$ symbol indicates the temperature of maximum isothermal compressibility ($\kappa_T^{\rm max}$-isobar) for liquid water along these isobars. (iv) The density profile of the critical vapor nucleus along the $P = -0.04$ GPa isobar at three different temperatures is shown. (B) The same for the CPF scenario of Model $1$ ($\epsilon_{\rm HB} = -16$ kJ/mol) is shown. Here, the $\blacktriangle$ symbol denotes the LLC temperature for pressures studied. We observe a distinct minimum and maximum in the $T$-dependent behavior of $\beta \Delta \Omega_{\rm v}^*$ under isobaric cooling. Notably, we find a pronounced non-classical wetting-mediated vapor nucleation near $T = 170$ K where the vapor nucleus is wetted by the intermediate LDL phase, as well as an Ostwald's step rule-like transition at temperatures below $170$ K (see also Fig.~\ref{fig:s3} in the Supplementary Material).}
    \label{fig:3}
\end{figure*}
\subsection{Non-classical vapor nucleation at negative pressures on isobaric cooling}
In this section, we have probed vapor nucleation along negative pressure isobars for the phase diagrams corresponding to different thermodynamic scenarios proposed to explain the anomalies of supercooled water. The results for Model $1$ are presented below and for Model $2$ are discussed in the Supplementary Material (see Fig.~\ref{fig:s4}). For Model $2$, at lower temperatures, the coexistence (LVC) density of the vapour phase becomes extremely low ($\sim 10^{-7}$ g/cc), and at such low densities, obtaining a converged nucleation barrier becomes challenging due to numerical inaccuracies.. Therefore, we have not reported the low-temperature results for Model $2$. 
\subsubsection{Two critical points (TCP) and critical point free (CPF) scenarios}
For the TCP scenario of Model $1$, the parameter $\epsilon_{\rm HB}$ is set to $-22$ kJ/mol. Figures~\ref{fig:3}A(i) and~\ref{fig:3}A(ii) present the phase diagram in the $P-T$ and $T- \rho$ planes, respectively.  In Fig.~\ref{fig:3}A(iii), we show the scaled nucleation barrier ($\beta \Delta \Omega_{\rm v}^*$) along three isobaric cooling paths (as indicated in Figs.~\ref{fig:3}A(i) and~\ref{fig:3}A(ii)). To calculate $\Delta \Omega_{\rm v}^*$ using the CDFT, we begin by obtaining the equilibrium (unstable) density profile, $\rho(r)$, for the critical vapor nucleus (see Fig.~\ref{fig:3}A(iv)). The CDFT allows direct computation of the equilibrium density profile without relying on a capillary approximation~\cite{oxtoby_2000, oxtoby_1996, oxtoby_1998, oxtoby_1994, ssb_2013, singh_2014}. This involves minimizing the grand potential of the inhomogeneous system:
\begin{equation}
    \Omega[\rho(\mathbf{r})] = \int d\mathbf{r} [f(\rho(\mathbf{r})) - \mu \rho(\mathbf{r})] + \int d\mathbf{r} \frac{1}{2} [K_\rho (\bigtriangledown \rho(\mathbf{r}))^2],
    \label{E7}
\end{equation} 
with respect to the density profile $\rho(\mathbf{r})$ by solving the Euler-Lagrange equation, $\delta \Omega [\rho (\mathbf{r})] / \delta \rho (\mathbf{r}) = 0$, with appropriate boundary conditions; $d\rho/dr=0$ at $r=0$, and $\rho(r)=\rho_l$ at $r \to \infty$). 
The square gradient term in the grand potential functional (Eq.~\ref{E7}) accounts for non-local effects resulting from density inhomogeneity. 

The nucleation barrier $\Delta \Omega_{\rm v}^*$ represents the extra free energy required (relative to the metastable homogeneous liquid phase) for forming the unstable equilibrium density profile of the critical cluster and is given by $\Delta \Omega_{\rm v}^* = \Omega[\rho(\mathbf{r})] - \Omega[\rho_l]$. We observed that $\beta \Delta \Omega_{\rm v}^*$ increases on isobaric cooling. However, near the Widom line (or, the locus of $\kappa_T^{\rm max}$ along isobars), there is an inflection point in the $T$-dependent behavior of $\beta\Delta\Omega_{\rm v}^*$ (see Fig.~\ref{fig:3}A(iii)). A similar $T$-dependent crossover behavior is observed in ice nucleation from supercooled water near the Widom line~\cite{singh_2014,poole_2015}. This crossover can be attributed to enhanced density fluctuations caused by the flattening of the free energy surface along the density near the Widom line. This flattening reduces surface tension by diffusing the interface. As a result, although the metastability of liquid water relative to the vapor phase gradually decreases during isobaric cooling, the reduction in surface tension near the Widom line offsets the effect of the decrease in metastability on the nucleation barrier. This interplay leads to a weaker $T$-dependence (or more precisely, a slight decrease) in $\beta\Delta\Omega_{\rm v}^*$ near the Widom line. This observation highlights the intricate relationship between modifications in the free energy landscape and nucleation kinetics. The TCP scenario for Model $1$ has already been explored in great detail in our previous work~\cite{singh_jcp_2022} and shows a similar $T$-dependent behavior of $\beta\Delta\Omega_{\rm v}^*$ along isobars.
\begin{figure*}[t]
    \centering
    \includegraphics[width = 0.94\textwidth]{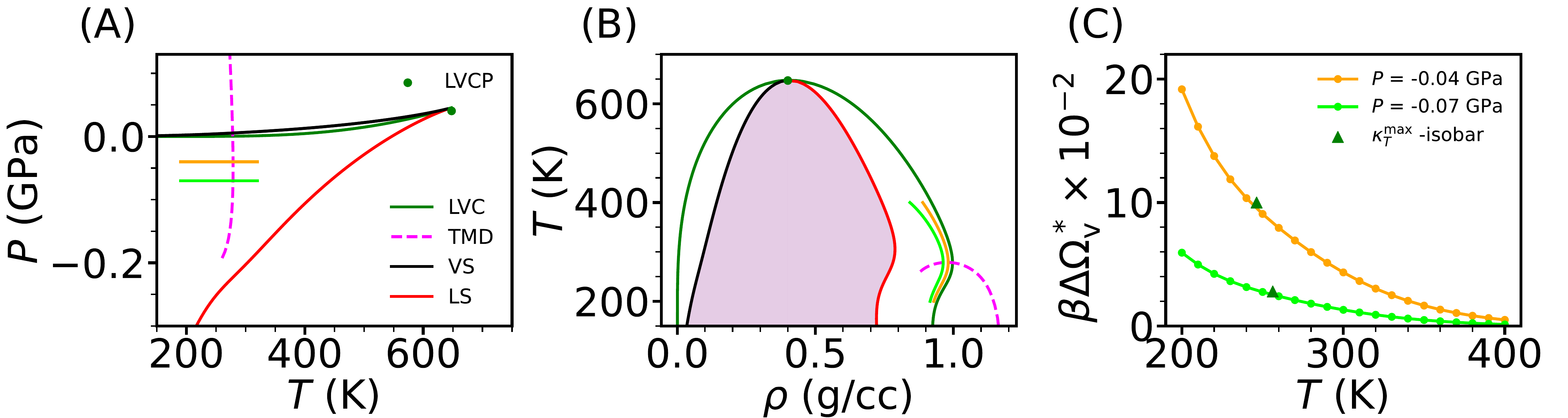}
    \caption{Isobaric cooling pathways at negative pressures are shown in the $P-T$ (A) and $T-\rho$ (B) planes for the SF scenario of Model $2$ ($\epsilon_{\rm HB} = -23$ kJ/mol; see Ref.~\emph{\citenum{pablo_jcp_1999}}). (C) The $T$-dependent scaled vapor nucleation barrier ($\beta \Delta \Omega_{\rm v}^*$) along two isobars is shown. The $\beta \Delta \Omega_{\rm v}^*$ increases monotonically as temperature decreases.}
   \label{fig:4}
\end{figure*}

For the CPF scenario ($\epsilon_{\rm HB} = -16$ kJ/mol for Model $1$), the LDL-HDL coexistence curve does not terminate at the anticipated critical point (see Section~\ref{model_details} and Fig.~\ref{fig:s1} in the Supplementary Material for details). Instead, the LDL branch is constrained within a dome-shaped LDL spinodal (LDLS). Here, the LS re-enters the positive pressure region at lower temperatures (see Figs.~\ref{fig:3}B(i) and~\ref{fig:3}B(ii)). On isobaric cooling, $\beta \Delta\Omega_{\rm v}^*$ initially increases and at lower temperatures near the LDL-HDL coexistence, it exhibits a strong non-monotonic behavior with a distinct minimum and maximum. Upon further cooling, $\beta \Delta\Omega_{\rm v}^*$ increases sharply. In the $P-T$ phase diagram (Fig.~\ref{fig:3}B(i)), the isobars seem to intersect with the LS at lower temperatures, so it is reasonable to speculate that at lower temperatures, near the LS, the liquid should spontaneously transition to the vapor phase. However, our results show a different outcome: instead of spontaneous vaporization, we observe a rise in $\beta \Delta \Omega_{\rm v}^*$ when the system is cooled below the LLC (Fig.\ref{fig:3}B(iii)). This is because, although the isobars appear to approach the LS in the $P-T$ plane, they actually transition to the region inside the LDLS dome, away from the LS, upon further cooling (see the isobars in the $T-\rho$ plane; Fig.~\ref{fig:3}B(ii)). This results in an increase in the nucleation barrier as cooling continues.

We further probed the nature of the density profile of the critical vapor nucleus near the LLC (see Fig.~\ref{fig:3}B(iv) for $P = -0.04$ GPa isobar). At lower temperatures, near the LLC, we observe that several non-classical effects, such as wetting-mediated transitions and an Ostwald step rule-like mechanism~\cite{Ostwald_1, Ostwald_2}, shape the $T$-dependent behavior of $\beta \Delta\Omega_{\rm v}^*$. At higher temperatures (e.g., $T = 190$ K), the vapor phase grows within the metastable HDL following a pathway consistent with the CNT. However, on lowering the temperature to around $T = 170$ K, we observe a non-classical wetting-mediated pathway, in which the critical vapor nucleus is wetted by the intermediate LDL phase (see Fig.~\ref{fig:3}B(iv), and schematic representations in Fig.~\ref{fig:s3} of the Supplementary Material). Upon further lowing the temperature, we find that the HDL transitions to the intermediate LDL phase, and vapor nucleation subsequently occurs within the LDL. This mechanism aligns with Ostwald's step rule, which posits that the phase that forms out of a metastable state may not be the thermodynamically most stable phase, but rather the one closest in stability to the parent phase~\cite{Ostwald_1}. Notably, the nucleation barrier minimum ($\beta \Delta\Omega_{\rm v}^{\rm min}$) is situated very close to the LLC, showing a more pronounced decrease than in the TCP scenario. These non-classical transitions (wetting-mediated and Ostwald step rule-like transitions) highlight a key distinction between the TCP and CPF scenarios --- in addition to the observation that the $\beta \Delta \Omega_{\rm v}^*$ for the CPF scenario displays a distinct minimum, whereas the TCP scenario shows only a slight deviation without a pronounced minimum. It is worth noting that non-classical crystal nucleation in complex systems are well-documented in theoretical~\cite{oxtoby_1998, ssb_2013, singh_2014}, computational~\cite{frenkel_science, moore, kawasaki, tanaka_soft_2012} and experimental studies (e.g., see Ref.~\emph{\citenum{chung}}). However, observations of non-classical nucleation involving the vapor phase remain rare\cite{triple_1, triple_2}. 
\begin{figure*}
    \centering
    \includegraphics[width=.83\textwidth]{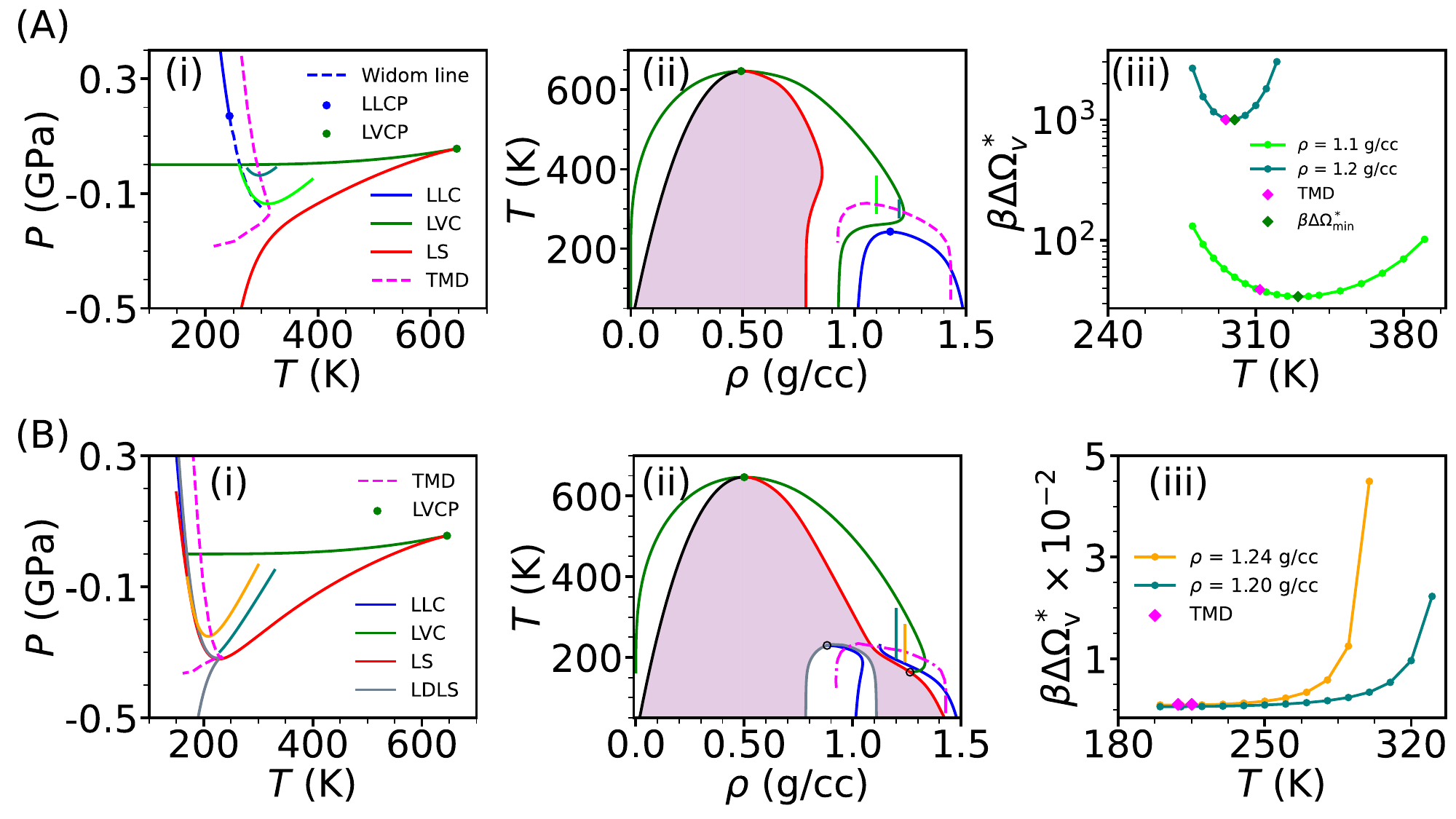}
    \caption{(A) Isochoric cooling pathways in the $P-T$ (i) and $T-\rho$ (ii) planes are shown for the TCP scenario of Model $1$. (iii) We show the $T$-dependent scaled vapor nucleation barrier ($\beta \Delta \Omega_{\rm v}^*$) along two isochores, where $\beta \Delta \Omega_{\rm v}^*$ exhibits non-monotonic behavior. The temperature of the minimum vapor nucleation barrier lies in the close vicinity of the TMD where the liquid water is maximally stretched. (B) The same is shown for the CPF scenario of Model $1$. Here, both the isochores exhibit a monotonic $T$-dependent $\beta \Delta \Omega_{\rm v}^*$. Pink diamonds mark the TMD along each isochore.}
    \label{fig:5}
\end{figure*}   
\subsubsection{Singularity free (SF) scenario}
We used Model $2$ to study the SF scenario as the parameters required to reproduce this behavior have already been documented by Truskett et al.~\cite{pablo_jcp_1999}. For this case, the $\epsilon_{\rm HB}$ is set to $-23$ kJ/mol, and other model parameters (such as, $r_i$, $r_o$ and $\phi^*$) differ from those used for the TCP scenario (see Ref.~\emph{\citenum{pablo_jcp_1999}} for details). To our knowledge, the SF scenario has not yet been reproduced within the framework of Model $1$. We again computed here the $T$-dependent vapor nucleation barrier along $P = -0.04$ and $-0.07$ GPa isobars (see Figs.~\ref{fig:4}A and~\ref{fig:4}B). The results reveal a monotonic increase in the nucleation barrier during isobaric cooling (Fig.~\ref{fig:4}C). This behavior is consistent with the characteristics of the underlying free energy surface. Unlike other cases, the SF scenario lacks singularities such as the LLCP or a line of stability at lower temperatures, which could soften the free energy surface at lower temperatures. Consequently, as temperature decreases, the isobars shift away from the LS in the $P-T$ plane, leading to an increase in $\beta \Delta\Omega_{\rm v}^*$ (Fig.\ref{fig:4}A). We further note that, although the $T$-dependent $\kappa_T$ along isobars exhibits a maximum at low temperatures, this increase in $\kappa_T$ is considerably less pronounced than in the TCP scenario (see Fig.~\ref{fig:s5} in the Supplementary Material). Therefore, in the SF scenario, the influence of the enhanced density fluctuations (or the maximum in $\kappa_T$) on the interfacial free energy --- and subsequently, on $\beta \Delta\Omega_{\rm v}^*$ --- is minimal. 
\subsection{Vapor nucleation barrier at negative pressure on isochoric cooling}
We further investigated the vapor nucleation barrier at negative pressures during isochoric cooling for the TCP and CPF scenarios in Model $1$. Two isochoric cooling paths were chosen (see Figs.~\ref{fig:5}A(i-ii) and~\ref{fig:5}B(i-ii) for TCP and CPF, respectively), along which we computed $\beta \Delta \Omega^*_{\rm v}$ for both scenarios. In the TCP scenario, $\beta \Delta \Omega^*_{\rm v}$ exhibits a non-monotonic temperature dependence during isochoric cooling (see Fig.~\ref{fig:5}A(iii)), where it initially decreases with temperature, reaches a minimum, and then increases as the temperature is further reduced. This behavior is consistent with recent theoretical findings on Model $2$~\cite{singh2023anomalous} and experimental studies involving low-density isochores~\cite{frederic_natphys_2013}. Notably, the temperature at which the nucleation barrier reaches its minimum does not coincide with the temperature of maximum metastability (or, the TMD), as is also observed in Model $2$~\cite{singh2023anomalous} and experiments on stretched water~\cite{frederic_natphys_2013}. In our previous work on Model $2$, we attributed this non-monotonic behavior of $\beta \Delta \Omega^*_{\rm v}$ during isochoric cooling to the combined effects of the TMD --- where liquid water reaches its maximum metastability --- and the monotonic temperature dependence of $\gamma_{\rm vl}$. We anticipate that this explanation applies similarly to Model $1$ as well.

In Fig.~\ref{fig:5}B(iii), we present the $T$-dependent behavior of $\beta \Delta \Omega^*_{\rm v}$ during isochoric cooling for the CPF scenario. We examined two isochores, $1.20$ g/cc and $1.24$ g/cc. Unlike the TCP scenario, here we observe a monotonic decrease in the vapor nucleation barrier upon cooling, even past the TMD line. This behavior arises because, in the CPF scenario, the LS line extends to higher densities, and therefore, the TMD line lies in close proximity to the LS (see Fig.~\ref{fig:5}B(ii)). As a result, $\beta \Delta \Omega_{\rm v}^*$ for isochores with similar densities ($\rho \sim 1.0 -1.2$ g/cc) does not exhibit non-monotonic behavior. This difference in the $T$-dependent behavior of $\beta \Delta \Omega_{\rm v}^*$ between the TCP and CPF scenarios offers a potential method for distinguishing between them. Furthermore, in the SF scenario, the $T$-dependent behavior of $\beta \Delta \Omega_{\rm v}^*$ along isochores is quite similar to that observed in the TCP scenario (see Fig.~\ref{fig:s2} in the Supplementary Material).
\section{Conclusions}\label{conclusions}
In this study, we aimed to investigate whether different theoretical scenarios proposed to explain the anomalies of supercooled water can be distinguished by examining the vapor nucleation kinetics and mechanisms from liquid water at negative pressures. To achieve this, we applied a classical density functional theory combined with realistic water models to explore how thermodynamic characteristics across these scenarios influence vapor nucleation kinetics under negative pressure conditions. We first analyzed the surface free energy between coexisting liquid and vapor phases, $\gamma_{\rm lv}$, in various scenarios. Notably, in all cases --- except the singularity-free scenario --- we found that the $\gamma_{\rm lv}$ along the LVC shows a non-monotonic dependence on $T$ and the temperature at which $\gamma_{\rm lv}$ reaches its minimum coincides with the temperature of maximum isothermal compressibility along the LVC line. However, the temperature at which $\gamma_{\rm lv}$ reaches its maximum did not correspond to the temperature of maximum equilibrium density difference between the coexisting phases along the LVC.

We then investigated the nucleation barrier during both isobaric and isochoric cooling for various scenarios. Our findings reveal notable differences in nucleation kinetics and mechanisms across these scenarios. In the TCP scenario, the nucleation barrier $\beta \Delta \Omega_{\rm v}^*$ increases steadily during isobaric cooling, with a slight decrease near the Widom line, followed by a subsequent rise as the temperature decreases further. In contrast, the nucleation barrier in the SF scenario increases monotonically throughout isobaric cooling.  Furthermore, in the CPF scenario, we report non-classical --- wetting mediated (where the emerging vapor  nucleus is wetted by the intermediate LDL phase), and Ostwald's step rule (where the HDL first transitions to the intermediate LDL phase rather than directly to the globally stable vapor phase) --- scenarios at low temperatures. Additionally, the analysis of vapor nucleation kinetics during isochoric cooling revealed significant differences between the TCP and CPF scenarios. Specifically, the $T$-dependent behavior of $\beta \Delta \Omega_{\rm v}^*$ in the CPF scenario exhibited distinct maxima and minima during isobaric cooling, with the minima occurring near the LDL-HDL coexistence line in the negative pressure region. We also investigated the behavior of $\beta \Delta \Omega_{\rm v}^*$ during isochoric cooling for Model $1$. In the TCP scenario, we consistently observed that the nucleation barrier exhibited a non-monotonic $T$-dependence. However, in the CPF scenario, the same isochores showed a monotonic decrease as the temperature decreased. 

Finally, it is important to note the limitations of our approach. First, the SLC and CPF scenarios are intertwined (the LS retraces back towards positive pressure for the CPF scenario) for both theoretical models (see Fig.~\ref{fig:1}(B) above and Fig.~\ref{fig:s1}(B) in the Supplementary Material), preventing us from clearly distinguishing between them in our study. Additionally, in Model $1$, the correlation length $K_\rho$ in the square gradient free energy functional is assumed to be independent of temperature (see Eqs.~\ref{E6} and~\ref{E7}), which could affect the quantitative accuracy of the vapor nucleation barrier. Nonetheless, this assumption is not expected to significantly alter the general $T$-dependent behavior of $\beta \Delta \Omega_{\rm v}^*$ along both isobars and isochores.  
  
\begin{acknowledgments}
R.S.S. acknowledges financial support from DST-SERB (Grant No. CRG/2023/002975). Y. S. acknowledges financial support from IISER Tirupati. M.S. acknowledges financial support from DST-SERB (Grant No. SRG/2020/001385). 
\end{acknowledgments}

\bibliographystyle{apsrev4-2}
\bibliography{water}
\balance

\pagebreak
\widetext
\begin{center}
\textbf{\large Supplementary Material}
\end{center}
\setcounter{equation}{0}
\setcounter{figure}{0}
\setcounter{table}{0}
\setcounter{page}{1}
\setcounter{section}{0}
\makeatletter
\renewcommand{\theequation}{S\arabic{equation}}
\renewcommand{\thefigure}{S\arabic{figure}}
\begin{figure*}[h]
    \centering
    \includegraphics[width = .6\textwidth]{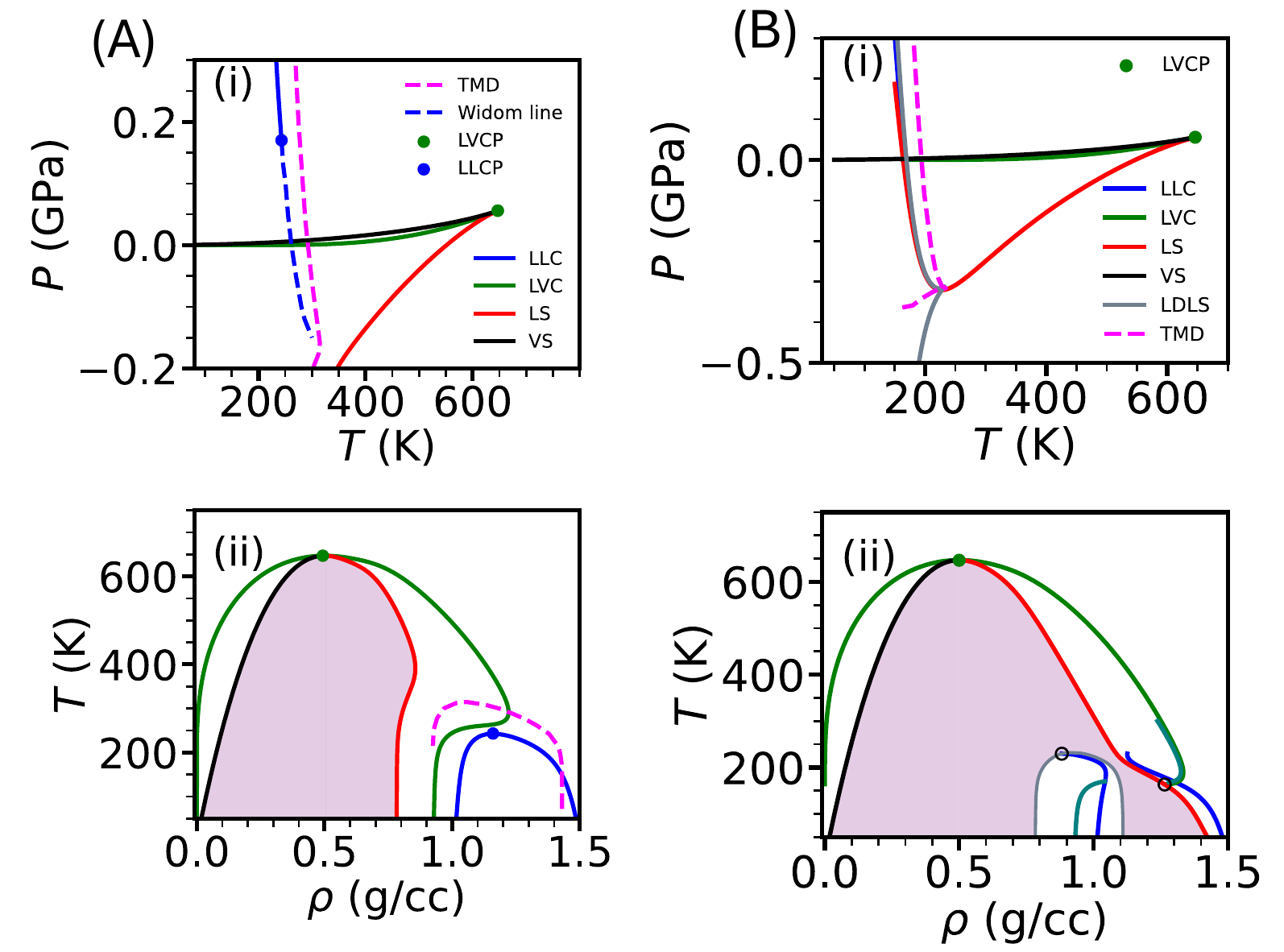}
    \caption{We show the phase digram of Model $1$ in the $P-T$ (i) and $T-\rho$ (ii) planes for two different H-bond strengths, $\epsilon_{\rm HB} = -22$ kJ/mol and $-16$ kJ/mol. As observed in Model $2$, these conditions produce two distinct scenarios --- TCP (A) and CPF (B)~\cite{poole1994effect}. To the best of our knowledge, the SF scenario has not yet been reproduced within the framework of Model $1$.}
    \label{fig:s1}
\end{figure*}

\begin{figure*}[h]
    \centering
    \includegraphics[width = .82\textwidth]{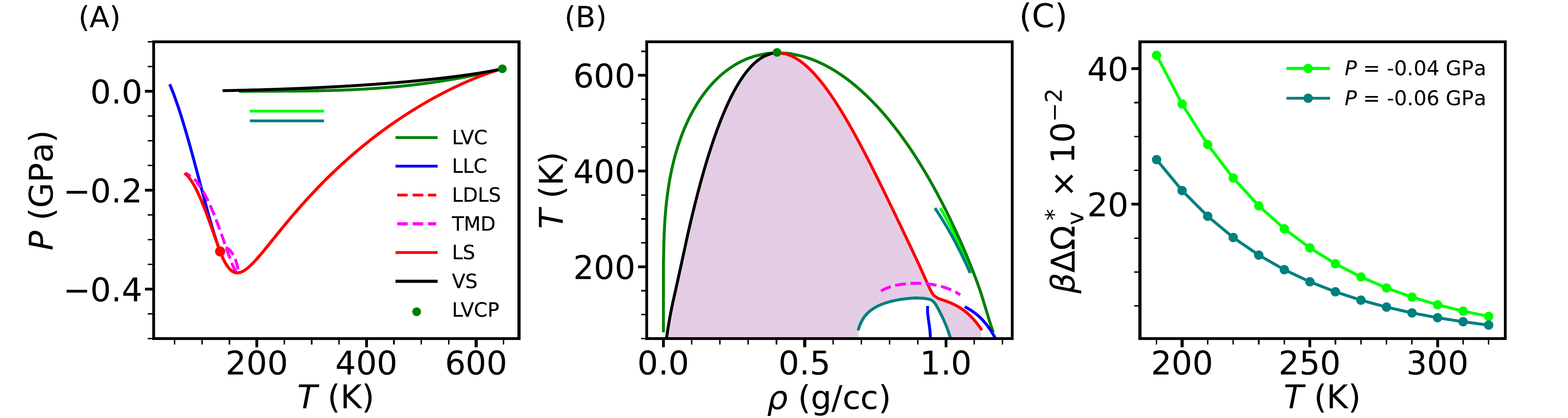}
    \caption{The phase diagram for the CPF scenario (Model $2)$ is shown in the $P-T$ (A), and $T-\rho$ (B) planes, along with the scaled vapor nucleation barrier ($\beta \Delta \Omega_{\rm v}^*$) along isobaric cooling paths (C). The filled red circle marks the speedy point. At lower temperatures, the (liquid-vapor) coexistence density of the vapor phase becomes very low ($\sim 10^{-7}$ g/cc), making it challenging to obtain a converged nucleation barrier due to numerical errors. Therefore, we have not reported the results at these low temperatures for this model.}
    \label{fig:s4}
\end{figure*}

\begin{figure*}[h]
    \centering
    \includegraphics[width = .75\textwidth]{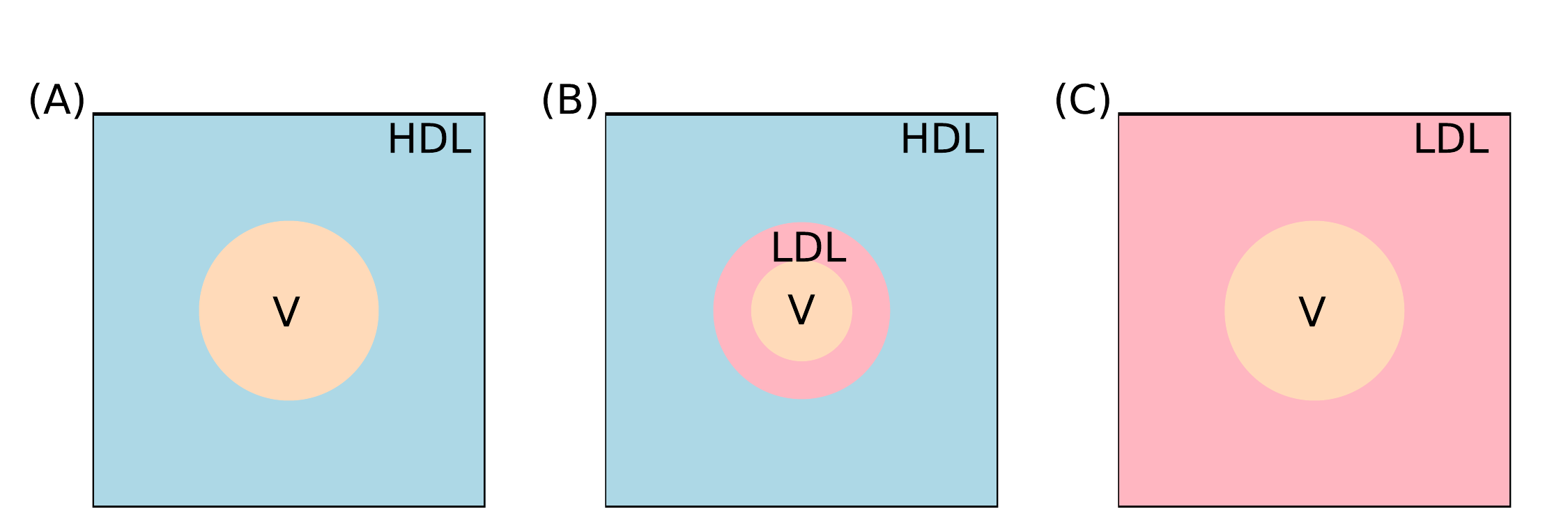}
    \caption{A schematic representation of the growing vapor nucleus (indicated by ``V") within the metastable liquid water phase is shown based on the density profiles in Fig.~\ref{fig:3}B(iv). (A) At $T = 190$ K, the vapor nucleates and grows within the HDL phase. (B) At $T = 170$ K, the vapor nucleus forms inside the HDL phase, however, the vapor-liquid interface is wetted by the intermediate density LDL phase. (C) At $T = 168$ K, the vapor grows within the metastable LDL phase. It is interesting to note here a crossover in nucleation mechanism from classical to non-classical. Specifically, a wetting-mediated and Ostwald-step rule like (the metastable HDL phase first transitions into the intermediate LDL phase rather than directly to the globally stable vapor phase) pathways emerge.}
    \label{fig:s3}
\end{figure*}

\begin{figure*}[h]
    \centering
    \includegraphics[width = .5\textwidth]{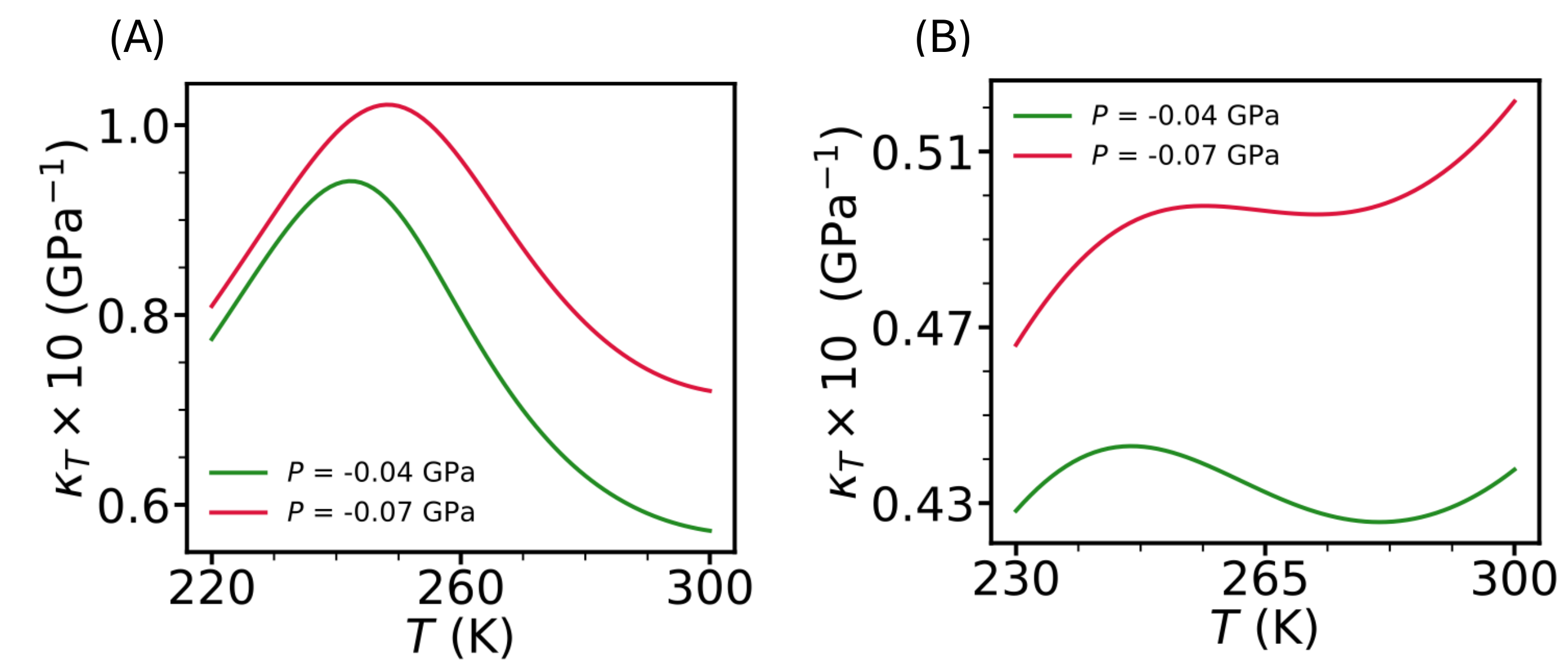}
    \caption{The $T$-dependent isothermal compressibility ($\kappa_T$) is shown along two isobars for the TCP (A) and SF (B) scenarios of Model $2$.}
    \label{fig:s5}
\end{figure*}

\begin{figure*}[h]
    \centering
    \includegraphics[width = .84\textwidth]{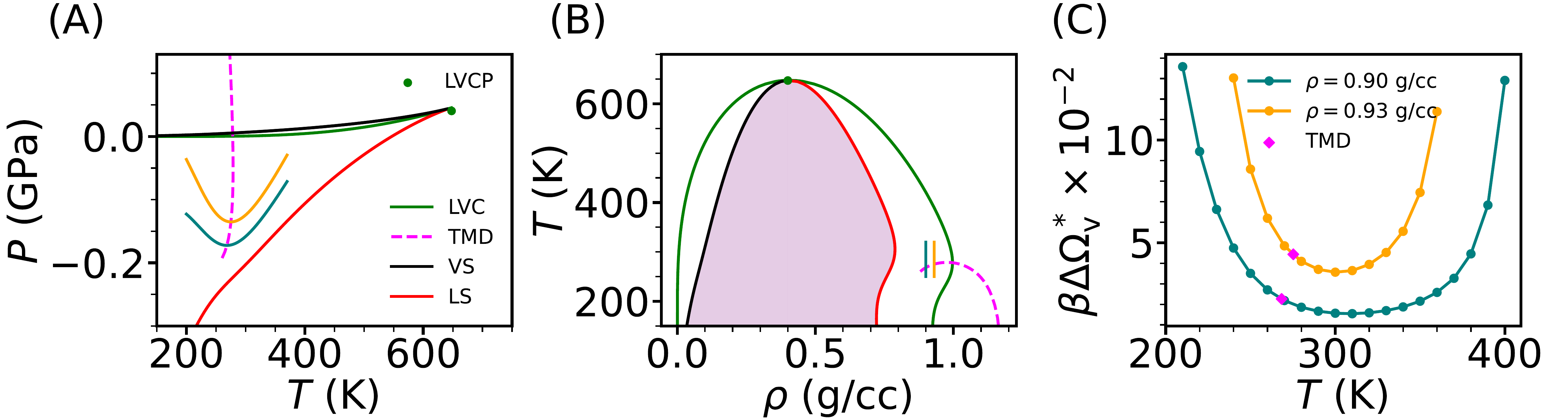}
    \caption{Two distinct isochoric cooling paths are shown in the $P-T$ (A) and $T-\rho$ (B) planes for the SF scenario of Model $2$. (C) The $T$-dependent scaled vapor nucleation barrier ($\beta \Delta \Omega_{\rm v}^*$) is shown for the $0.90$ g/cc and $0.93$ g/cc isochores. We note that, the behavior of $\beta \Delta \Omega_{\rm v}^*$ during isochoric cooling in this scenario closely resembles that observed in the TCP scenario.}
    \label{fig:s2}
\end{figure*}


\end{document}